\documentclass[aps,pra,reprint,superscriptaddress]{revtex4-1}
\usepackage{epic,eepic}
\usepackage{graphicx}
\usepackage{dcolumn}
\usepackage{bm}
\usepackage{epstopdf}
\usepackage{amsmath}
\usepackage{amssymb}
\usepackage{latexsym}
\usepackage{revsymb}

\begin{document}



\title{Expansion of an ultracold Rydberg plasma}


\author{Gabriel T. Forest}
\affiliation{Department of Physics and Astronomy, Colby College, Waterville, ME 04901, USA}

\author{Yin Li}
\affiliation{Department of Physics and Astronomy, Colby College, Waterville, ME 04901, USA}

\author{Edwin D. Ward}
\affiliation{Department of Physics and Astronomy, Colby College, Waterville, ME 04901, USA}

\author{Anne L. Goodsell}
\affiliation{Department of Physics, Middlebury College, Middlebury, VT 05753, USA}

\author{Duncan A. Tate}
\email[]{duncan.tate@colby.edu}
\affiliation{Department of Physics and Astronomy, Colby College, Waterville, ME 04901, USA}


\date{\today}

\begin{abstract}
We report a systematic experimental and numerical study of the expansion of ultra-cold Rydberg plasmas. Specifically, we have measured the asymptotic expansion velocities, $v_0$, of ultra-cold neutral plasmas (UNPs) which evolve from cold, dense samples of Rydberg rubidium atoms using ion time-of-flight spectroscopy. From this, we have obtained values for the effective initial plasma electron temperature, $T_{e,0} = m_{ion} v_0^2/k_B$ (where $m_{ion}$ is the Rb$^+$ ion mass), as a function of the original Rydberg atom density and binding energy, $E_{b,i}$. We have also simulated numerically the interaction of UNPs with a large reservoir of Rydberg atoms to obtain data to compare with our experimental results. We find that for Rydberg atom densities in the range $10^7 - 10^9$ cm$^{-3}$, for states with principal quantum number $n > 40$, $T_{e,0}$ is insensitive to the initial ionization mechanism which seeds the plasma. In addition, the quantity $k_B \, T_{e,0}$ is strongly correlated with the fraction of atoms which ionize, and is in the range $0.6 \times |E_{b,i}| \lesssim k_BT_{e,0} \lesssim 2.5 \times |E_{b,i}|$. On the other hand, plasmas from Rydberg samples with $n \lesssim 40$ evolve with no significant additional ionization of the remaining atoms once a threshold number of ions has been established. The dominant interaction between the plasma electrons and the Rydberg atoms is one in which the atoms are deexcited, a heating process for electrons that competes with adiabatic cooling to establish an equilibrium where $T_{e,0}$ is determined by their Coulomb coupling parameter, $\Gamma_e \sim 0.01$. 
\end{abstract}

\pacs{}

\maketitle

\section{{INTRODUCTION}\label{intro}}
The behavior and properties of ultra-cold neutral plasmas (UNPs) made by direct photoionization of cold atoms in a magneto-optical trap (MOT), first discovered in 1999 \cite{kill99}, are now relatively well understood (see for instance Refs. \cite{kill07,lyon17}). Above the ionization threshold, $E_I$, conservation of linear momentum in the ionization process dictates that most of the excess photon energy, $\Delta E = h \nu - E_I$, goes to the electron. When the ionizing laser is tuned well above threshold, the initial electron temperature, $T_{e,0}$, is given by $\Delta E = \frac{3}{2} k_B T_{e,0}$. The asymptotic plasma expansion velocity of the plasma is given by 
\begin{equation}
v_0 = \sqrt{\frac{k_B (T_{e,0} + T_{ion,0})}{m_{ion}}}, \label{vZero}
\end{equation} 
where $m_{ion}$ is the ion mass. The initial ion temperature, $T_{ion,0}$, is determined largely by the temperature of the parent atoms in the MOT and is typically in the range $0.1 - 10$ mK for UNPs made from noble gas atoms, alkalis, or alkaline earths. However, a number of mechanisms rapidly heat both electrons and ions. Specifically, close to threshold, three body recombination (TBR) heats the electrons and results in minimum $T_{e,0}$ values in the range 30 - 50 K, and, at high density, threshold lowering (TL) will also affect $T_{e,0}$ \cite{hahn02}. (That is, these mechanisms cause $v_0$ to be higher than Eq. \ref{vZero} predicts, based on the $T_{e,0}$ determined by the excess energy of the ionizing photon.) Additionally, the ions are subject to disorder induced heating (DIH), which heats them up to $\sim 1$ K in the first few microseconds of the plasma evolution process at higher densities \cite{chen04,lyon17}, but is much less significant in UNPs created in the low density regime \cite{wils13,chen17}. As the plasma expands adiabatically on a time scale of order 10 - 100 $\mu$s, both the electron and ion temperatures fall below the initial values determined by $\Delta E$, TBR, TL, and DIH. Additionally, the Coulomb coupling parameter, $\Gamma_\alpha$, increases \cite{kill07}, where
\begin{equation}
\Gamma_\alpha = \frac {e^2}{4 \pi \epsilon_0 a_{\alpha} k_B T_\alpha}, \label{gamma}
\end{equation}
and $a_{\alpha}$ is the Wigner-Seitz radius for species $\alpha$ (which may be electrons, $e$, or ions, $ion$). As a consequence of the competition between adiabatic cooling and TBR, it has been shown that, for typical initial conditions, UNPs tend to equilibrate to $\Gamma_e \sim 0.1$ \cite{robx02}.

UNPs also evolve spontaneously from dense samples of cold Rydberg atoms. Such plasmas (herein termed Rydberg plasmas) are made from cold atoms in a MOT. They were first reported in Refs. \cite{rols98,rob00}, though a similar phenomenon in dense thermal samples in an atomic beam was observed much earlier \cite{vit82}. Additionally, Rydberg plasmas have also been created using translationally cool atoms and molecules in a supersonic beam \cite{morr08,schu16b}. For Rydberg plasmas created in a MOT, it has been shown that dipole interactions between cold (``frozen'') Rydberg atoms play a significant role in the initial ionization \cite{tann08,robx14}, and black body radiation (BBR) \cite{bet09c,spen82b}, and collisions with hot background Rydberg atoms also contribute \cite{li04}. Once a critical electron density threshold is achieved, an avalanche of electron-Rydberg collisions is initiated, and the plasma evolves mediated by the exchange of energy between the Rydberg atoms and the UNP. However, to our knowledge, there have been no extensive experimental investigations of the dependence of the expansion velocity of a Rydberg plasma on density and initial binding energy, $E_{b,i}$. 

The ionization processes which initiate Rydberg plasmas, and the effect of the UNP so created on the Rydberg state distribution, have been considered theoretically in a number of papers (see, for example, Refs. \cite{robx03,poh03}). However, these papers do not discuss any correlation of the plasma electron temperature with the changing state distributions due to the presence of the daughter UNP. There has also been a theoretical investigation of the feasibility of reaching the strongly coupled regime for the ions in UNPs ($\Gamma_{ion} \gtrsim 1$) using dipole-blockaded cold Rydberg samples \cite{bann13}, and several experimental studies using optical imaging of Rydberg plasmas of the critical processes during the avalanche regime \cite{rdsv13,sier14,mcqu13}. In particular, in Ref. \cite{rdsv13}, an electron temperature of $30 \pm 10$ K was found at the end of the avalanche for a UNP which evolved from $55s_{1/2}$ $^{87}$Rb atoms at densities $\lesssim 10^{11}$ cm$^{-3}$, while in Ref. \cite{sier14}, temperatures of 26.0 K and 48.7 K were found for UNPs from $^{87}$Rb samples in the $45s_{1/2}$ and $40d$ states, respectively. However, in both studies, these temperatures were found indirectly, from models that describe how the optical depth in absorption imaging of a particular hyperfine component of the $5s_{1/2} \rightarrow 5p_{3/2}$ transition is affected when the $5p_{3/2} \rightarrow n\ell$ transition is excited with narrow bandwidth laser pulses of duration 5 - 35 $\mu$s \cite{rdsv13} and 200 $\mu$s \cite{sier14}. Furthermore, these papers give no information on how the electron temperatures found depend on density, how the interaction of the plasma and the Rydberg atom reservoir affects the Rydberg atoms, and for each paper, only one or two states were investigated. Finally, there has been extensive work done on UNPs which evolve from translationally cold samples of Rydberg NO molecules in a supersonic beam by a group at University of British Columbia (see \cite{schu16b} and references therein). Such Rydberg plasmas have significantly more complex behaviors than do atomic UNPs due to additional dynamical pathways available to molecular systems.

This paper reports a systematic experimental and numerical study of the asymptotic expansion velocity of UNPs which evolve from cold Rydberg samples, as a function of the initial binding energy, and the initial Rydberg density. From $v_0$, we use Eq. \ref{vZero} to infer a value for $T_{e,0}$ for such plasmas. This ``effective initial electron temperature'' is a phenomenological parameter which is related to the mean kinetic energy of an ion's outward velocity late in the plasma evolution. Nevertheless, $T_{e,0}$ is a standard parameterization of the electron thermal energy early in the plasma evolution, and allows comparisons to be made with UNPs made by direct photoionization \cite{kill07}. First, we describe our experiments, and the critical results. Then, we discuss our numerical modeling of these systems and how they substantially replicate the experimental findings. The model is then used to gain an intuitive understanding of the interactions between the electrons and the Rydberg reservoir during the plasma evolution process.

\begin{figure}
\centerline{\resizebox{0.40\textwidth}{!}{\includegraphics{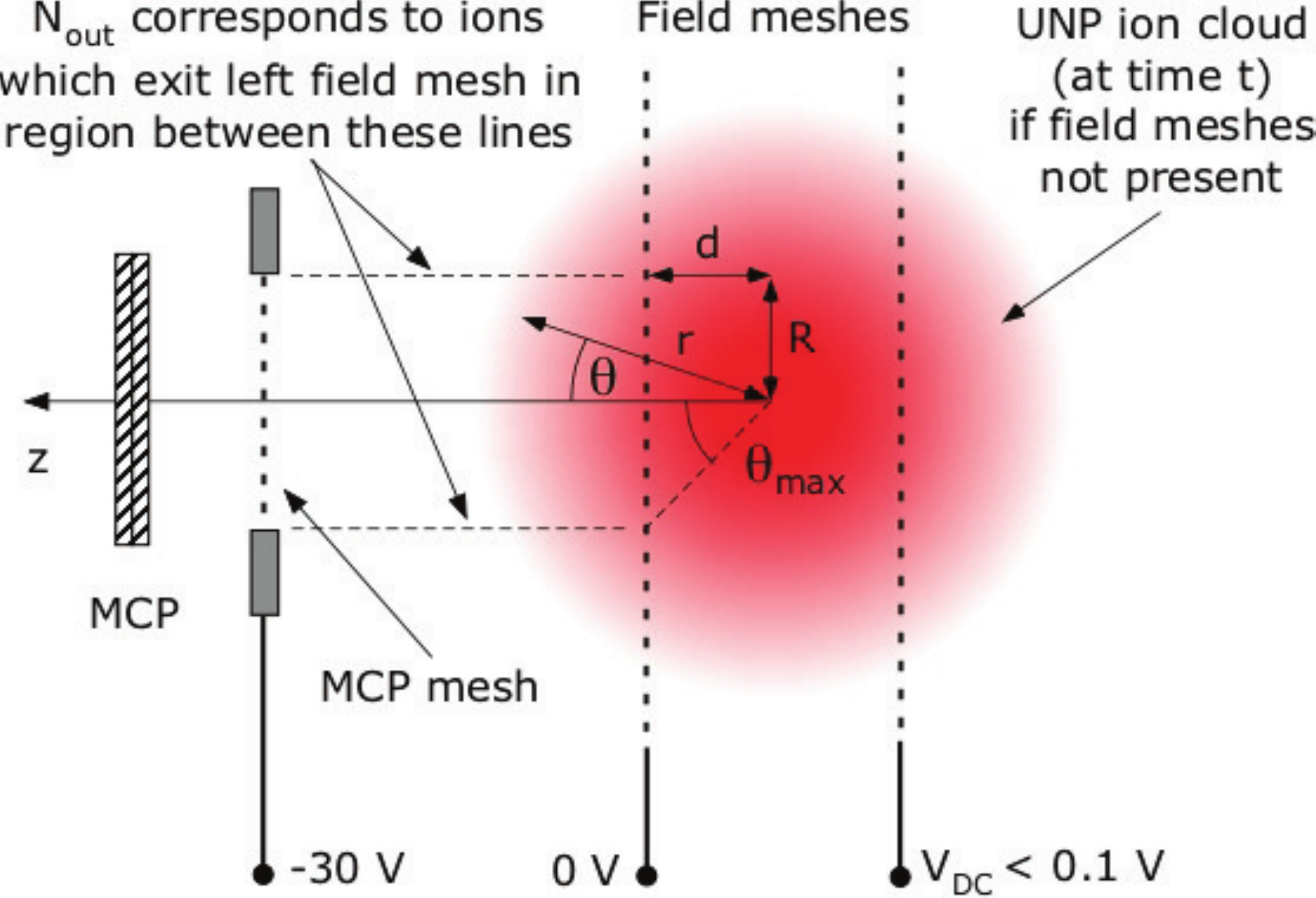}}}
\caption{Schematic of the field meshes and MCP used to obtain the plasma asymptotic expansion velocity from the ion TOF signal. The apparatus is cylindrically symmetric about the $z$-axis. Spherical coordinates $r$, $\theta$, and $\phi$ (the azimuthal angle) specify the location relative to the center of the plasma, $R$ is the effective acceptance radius of the MCP, and $d$ is half the field mesh spacing. The angle $\theta_{max} = \tan^{-1}(R/d)$ defines the effective maximum acceptance angle of the MCP. See text for details. $V_{DC}$ is a small voltage applied to null the effect of external fields in the interaction region between the field meshes.}
\label{ionTOF}
\end{figure}

\section{{APPARATUS}\label{expt}}

Our study concerns UNPs which evolve from cold $nd_j$ $^{85}$Rb Rydberg atoms ($24 \le n \le 120$). The effective initial electron temperature, $T_{e,0}$, of these UNPs is found by measuring their asymptotic expansion velocity, $v_0$, from ion time-of-flight (TOF) spectra. The Rydberg atoms are created from cold atoms in a MOT which has a maximum atom density of approximately $1 \times 10^{10}$ cm$^{-3}$ ($1/\sqrt{e}$ radius $\sigma_0 \approx 400 \ \mu$m) and atom temperature $\sim 100 \ \mu$K. The atoms are excited to the $nd_j$ states using a narrow-bandwidth pulsed laser system (NBPL) \cite{bran10}. Excitation of the cold atoms takes place between two parallel high-transparency copper meshes separated by 18.3 mm which may be biased to null out external fields, and we can also apply voltage pulses to selectively field ionize (SFI) Rydberg atoms. (We use SFI in this experiment only to remove atoms excited to Rydberg states from the trap in order to measure the Rydberg atom density as described below.) We monitor the plasma evolution, or the SFI signal, by either observing electrons or ions using a micro-channel plate detector (MCP). We achieve Rydberg densities in the range $1 \times 10^7 - 1 \times 10^9$ cm$^{-3}$, which we vary by changing the laser pulse energy. The number of atoms excited is monitored by measuring the 780 nm resonance fluorescence depletion when we apply an SFI pulse immediately after laser excitation \cite{han09}. (The SFI pulse has a magnitude which is significantly greater than the classical ionization threshold, $1/16{n^\ast}^4$ in atomic units, where $n^\ast$ is the effective principal quantum number of the Rydberg state.) The Rydberg atom densities have an absolute uncertainty of a factor of approximately 2, and a relative uncertainty of 20-30\%. The NBPL laser beam is unfocussed, with a diameter of $\approx 4$ mm. This is much larger than the size of our cold atom sample, whose diameter we measure by imaging the 780 nm fluorescence onto a linear diode array. To minimize systematic effects due to laser beam movement (which can affect the geometry of the interaction region relative to the MCP, with consequent impact on the parameters found from fitting the ion TOF signal), we used fixed apertures to define the NBPL beam axis, and optimized the position of the MOT atom cloud to this axis.

The field of ion TOF spectroscopy is well-developed, and exceedingly diverse. In the context of these experiments, the technique is applied to obtain the expansion velocity of a spherically symmetric ion cloud. A similar method has been used in molecular beam UNP experiments by the group at UBC, where the time-profile of the ion cloud is observed as a function of plasma evolution time, which is varied by simply moving the detection system longitudinally along the beam \cite{saqu11}. However, when the UNP center-of-mass is stationary, the situation is somewhat more complex, and the principle of the ion TOF technique we use is described in Ref. \cite{twed12}. After excitation, the cold Rydberg samples evolve to plasma over a period of $\le 10 \ \mu$s, which is negligible in comparison with the overall expansion time of the UNP (100 - 200 $\mu$s). The plasma slowly expands, and we use the MCP to detect plasma ions which exit the field-free interaction region between the meshes. Specifically, the ions we detect are those which leave the interaction region through the left-hand field mesh in Fig. \ref{ionTOF} and enter the cylindrical region centered on the $z$ axis whose cross section is defined by the MCP acceptance aperture. Mathematically, the number of ions in this volume of space, $N_{out}$, is 
\begin{widetext}
\begin{equation}
N_{out} = 2 \pi \int_{\theta=0}^{\theta_{max}} \, \int_{r=d/\cos \theta}^\infty \ \biggl [ \frac {N_{ion}}{(2 \pi \sigma^2)^{3/2}} e^{-\frac{r^2}{2\sigma^2}} \biggr ] \, r^2 \, \sin \theta \, dr \, d\theta, \label{nout}
\end{equation} 
\end{widetext}
where the term in square brackets inside the integral is the Gaussian ion density distribution, $\rho_{ion}(r,t)$ (the total ion number is $N_{ion}$), $r$ is the distance from the center of the plasma, and $t$ is the time since it was created. The plasma has a characteristic radius $\sigma(t) = \sqrt{\sigma^2_0 + v^2_0 \, t^2}$, and other quantities are defined in Fig. \ref{ionTOF}. (The equation has already been integrated with resect to $\phi$.)

Our MCP signal is proportional to the ion current, $\frac{dN_{out}}{dt}$, where $N_{out}$ is given by Eq. \ref{nout}, and we fit our ion TOF signal using this equation. We thus assume that the part of the UNP that lies between the field meshes is unaffected as the outermost ions exit this region and are accelerated towards the MCP. Since the mesh we use (Buckbee-Mears MC-4) has a transparency of 95\%, and the electric fields between the meshes, and between the left field mesh in Fig. \ref{ionTOF} and the MCP, have magnitudes $\lesssim 0.1$ V/cm and $\approx 10$ V/cm respectively, this assumption seems reasonable. Additionally, the use of similar meshes through which electrons or ions pass in imaging or TOF measurements on UNPs has been used extensively in other work, and no significant perturbations due to the meshes have been found (see, for example, Refs. \cite{morr08,zhan08a,twed12}). We assume the ions follow straight line paths parallel to the symmetry axis (the $z$-axis in Fig. \ref{ionTOF}) between the field mesh and the MCP mesh. While there may be some weak ion lensing effects in this region (discussed below), the maximum ion density here is of order $10^5$ cm$^{-3}$, too low for any significant Coulomb repulsion effects. The ions generally start to arrive at the MCP 30 - 40 $\mu$s after the NBPL pulse, and the MCP signal peaks between 80 and 100 $\mu$s, with an overall duration of $\lesssim 400 \ \mu$s. Since $\sigma_0 \ll v_0 \, t$ for all $t$ where the MCP signal is non-zero, we make the substitution $\sigma = v_0 \, t$. The time dependence of the signal predicted using Eq. \ref{nout} depends on only three free parameters: $v_0$, $N_{ion}$, and a geometric factor which depends on $\theta_{max}$. The MCP signal we detect depends additionally on an unknown but constant detection conversion factor (i.e., the current output for an ion flux of 1/s), which with $N_{ion}$ affects only the vertical scaling of the detected signal, and a time offset which reflects the flight time of the ions from where they exit the field meshes to the MCP itself. (For all our data, we subtracted off a background signal obtained with the MOT magnetic field turned off, but with the NBPL beam entering the interaction region.) Fitting our data using Eq. \ref{nout} and allowing the vertical scaling parameter to float, but with specified offset time and $\theta_{max}$ enabled us to extract a values for $v_0$. 

We use these $v_0$ values to obtain values for $T_{e,0}$ using Eq. \ref{vZero}, making the assumption that $T_{ion,0} \lesssim 1$ K, and is therefore negligible in comparison to $T_{e,0}$. We have carried out extensive calibration of this technique by using it to find $v_0$ values for UNPs made by photoionizing cold atoms in the limit where $T_{e,0}$ is well above the regime in which TBR is important ($T_{e,0} = 50 - 300$ K), and find $v_0$ to be in agreement with Eq. \ref{vZero}, if we ignore $T_{ion,0}$ and use $\Delta E = \frac{3}{2} k_B T_{e,0}$. The values of $T_{e,0}$ we obtain for Rydberg UNPs lie in the range 20 - 130 K, with an uncertainty of $\sqrt{(10 \ \textrm{K})^2 + (0.1 \times T_{e,0})^2}$. The ion TOF spectra exhibit small systematic differences from Eq. \ref{nout}, and this makes the $v_0$ values we obtained sensitive to the time offsets and MCP acceptance angles, which were kept constant in the fits. Our uncertainty estimate, found by fitting selected TOF spectra using a range of different time offsets and MCP acceptance angles, reflects the impact of three systematic effects which are not included in Eq. \ref{nout}. Specifically, the UNP density profile falls off more sharply than the Gaussian function assumed near the edges \cite{robx03,pohl04b}, and there are variations in the effective acceptance angle of the MCP due to ion lensing effects between the field mesh and the MCP. There is a mesh with voltage -30 V just in front of the MCP, while the field mesh nearest the MCP is grounded. This effectively forms a weak ion lensing system, and will result in a velocity-dependent effective aperture of the MCP. The hydrodynamic velocity of an ion, $\vec u$, is related to its position $\vec r$ relative to the center of the plasma by the parameter $\gamma(t)$, where 
\begin{equation}
\vec u(\vec r,t) = \gamma(t) \, \vec r = \frac{v_0^2}{\sigma(t)^2} \, t \, \vec r \label{ionvel}
\end{equation}
(see Ref. \cite{kill07}, Eqs. 24 - 26). Because $\vec u$ depends on $\vec r$ and $t$, there will be slight variations in the effective acceptance angle of the MCP over the course of the UNP evolution. Finally, during the course of the modeling described below in Section \ref{anal}, we found that UNPs which evolve from cold Rydberg samples take a relatively long time to reach a constant value for $v_0$. During the modeling, we found the $m_{ion}v_0^2/k_B$ values reached more than 95\% of their final values within 40 $\mu$s of the plasma creation, and this is less than the time-of-flight for the first ions we detect in our experiment. Nevertheless, since our fitting routine assumes $v_0$ is constant, the $v_0$ values we obtain will be subject to error from this source, too.

\begin{figure}
\centerline{\resizebox{0.50\textwidth}{!}{\includegraphics{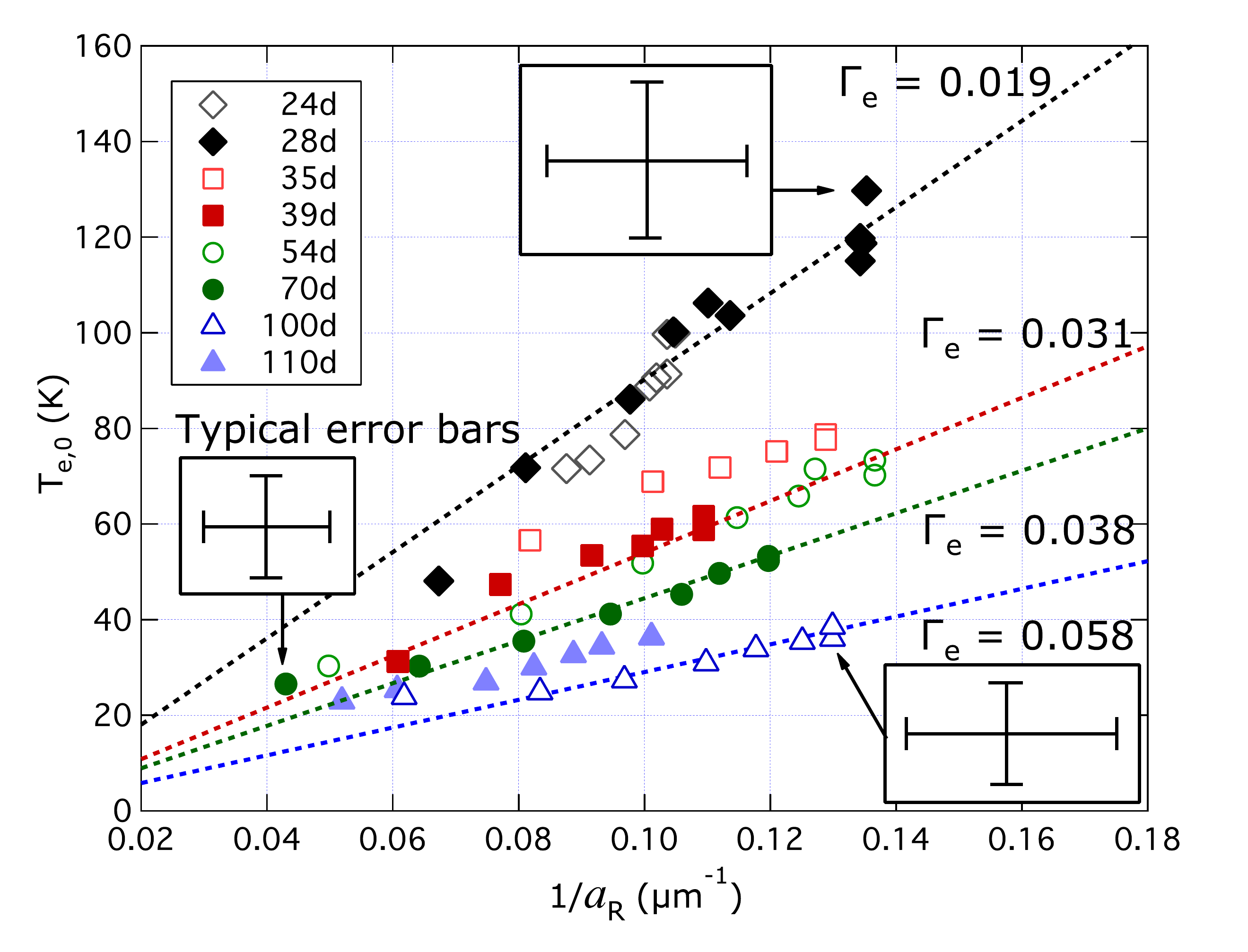}}}
\caption{Graph of $T_{e,0}$ versus $1/a_R$, where $a_R$ is the mean Rydberg atom spacing, for UNPs evolving from cold $nd$ Rb Rydberg samples in the range $24 \le n \le 110$. Within the experimental uncertainties, the data for a given Rydberg state are consistent with a straight line relationship (with $y$-intercepts of zero). The dashed lines are straight line fits that are constrained to have zero intercepts, and the corresponding $\Gamma_e$ values are given for $28d$, $39d$, $70d$, and $100d$, assuming that $a_e$, the Wigner-Seitz radius for electrons in the UNP, is equal to $a_R$. Typical error bars are as shown.}
\label{TeVsa}
\end{figure}

\section{{EXPERIMENTAL RESULTS}\label{exres}}

Typical results for $T_{e,0}$ are shown in Fig.\,\ref{TeVsa} as a function of the reciprocal of the Rydberg atom spacing, $1/a_R = (4 \pi \rho_{R,avg}/3)^{1/3}$, where $\rho_{R, avg}$ is the average Rydberg atom density, for different $nd$ states in the range $24 \le n \le 120$. The low $T_{e,0}$, low $1/a_R$ cutoff ($n = 24$) in the data is determined by the lowest density sample that would spontaneously evolve into a UNP, whereas the high $T_{e,0}$, high $1/a_R$ cutoff ($n = 120$) is determined by our maximum achievable density due to the declining oscillator strength of the $5p_{3/2} \rightarrow nd$ transition and the maximum pulse energy available from the NBPL. 

There are a number of interesting features in the data shown in Fig. \ref{TeVsa}. First, within the experimental uncertainties, the results for a single Rydberg state fall on a straight line whose $y$-intercept is zero. The data shown in Fig. \ref{TeVsa} therefore provide strong circumstantial evidence that the plasmas which form from a particular $nd$ state have approximately constant initial $\Gamma_e$ values, regardless of density. Specifically, if we rewrite Eq. \ref{gamma} for electrons, we see that
\begin{equation}
T_e = \biggl (\frac {e^2}{4 \pi \epsilon_0 k_B} \, \frac{1}{\Gamma_e} \biggr ) \, \frac{1}{a_{e}} = \frac{16.7 \, \textrm{K}\mu \textrm{m}}{\Gamma_e} \, \frac{1}{a_{e}}. \label{gammamod}
\end{equation}
Hence, for constant $\Gamma_e$, a plot of $T_{e}$ versus $1/a_e$ will be a straight line. The data in Fig. \ref{TeVsa} plot $T_{e,0}$ versus $1/a_R$, and the relationship between the mean Rydberg atom spacing and the Wigner-Seitz radius for the electrons depends on the fraction of Rydberg atoms which ionize, $f$, as $a_e/a_R = f^{-1/3}$. However, the $f^{-1/3}$ scaling makes $a_e/a_R$ relatively insensitive to the ionization fraction: a variation of $f$ from 0.1 to 0.8 changes $a_e/a_R$ by only a factor of 2. If we assume that $f = 1 \Rightarrow a_e = a_R$, we find that the values of $\Gamma_e$ vary from 0.02 ($28d$) to approximately 0.06 ($100d$). Lines of constant $\Gamma_e$ using this assumption are shown in Fig. \ref{TeVsa} for the $28d$, $39d$, $70d$, and $100d$ states. (Note that these $\Gamma_e$ values are characteristic of times early in the plasma evolution, found using the $T_{e,0}$ values inferred from the asymptotic $v_0$ values.)

The second feature apparent in Fig. \ref{TeVsa} is that $\Gamma_e$ generally increases as the magnitude of the initial binding energy, $E_{b,i}$, decreases, where 
\begin{equation}
E_{b,i} = -\frac{e^2}{4 \pi \epsilon_0} \frac {1}{{2 a_0 n^\ast}^2}, \label{be}
\end{equation}
in which $a_0$ is the Bohr radius and $n^\ast$ is the effective principal quantum number of the initial $nd$ state, $n^\ast \approx n - 1.35$. While there is some scatter from a monotonic relationship between $T_{e,0}$ and $|E_{b,i}|$ for fixed $1/a_R$, the scatter between close $nd$ states is within our experimental uncertainties. The data shown in Fig. \ref{TeVsa} indicate a significant correlation between $T_{e,0}$ and the initial Rydberg state binding energy, and this suggests that it would be useful to plot the data using the scaled quantities $\tilde T = k_B T_{e,0}/|E_{b,i}|$ and $\tilde a_e = a_e/2{n^\ast}^2a_0$. Using this scaling, it can be seen that Eq. \ref{gamma} for electrons can be expressed as
\begin{equation}
\Gamma_e = \frac {1}{\tilde a_e \, \tilde T}, \label{gammaScaled}
\end{equation}
and thus for constant $\Gamma_e$, one expects $\tilde T \propto 1/\tilde a_e$. 

To plot our data using Eq. \ref{gammaScaled} would require values for $a_e$, which in turn would need accurate measurements of $f$. In principle, $f$ can be measured at a particular evolution time by applying a sufficiently large negative-voltage SFI pulse and observing the electrons liberated as the UNP is quenched by the leading edge of the pulse and those which arrive later as the SFI pulse field ionizes successively more deeply bound Rydberg states. However, it is hard to do this with any degree of precision unless an accurate measurement is made of the number of free plasma electrons which escape before the SFI pulse is applied, and there is also significant uncertainty due to migration of Rydberg population to deeply bound states which cannot be field ionized. Additionally, $f$ changes during the course of the plasma evolution, and without extensive modeling, choosing the appropriate time to evaluate $f$ would also introduce significant uncertainty. As a consequence of these considerations, we opted to use numerical modeling to obtain values for $f$, $T_{e,0}$, and other related plasma parameters, as functions of plasma evolution time. From this, we standardized on a specific time during the plasma evolution, $t = 40 \ \mu$s, at which to take a snapshot of the evolution that we then compared against the experimental results. The numerical analysis is described below in Section \ref{anal}.

\begin{figure}
\centerline{\resizebox{0.50\textwidth}{!}{\includegraphics{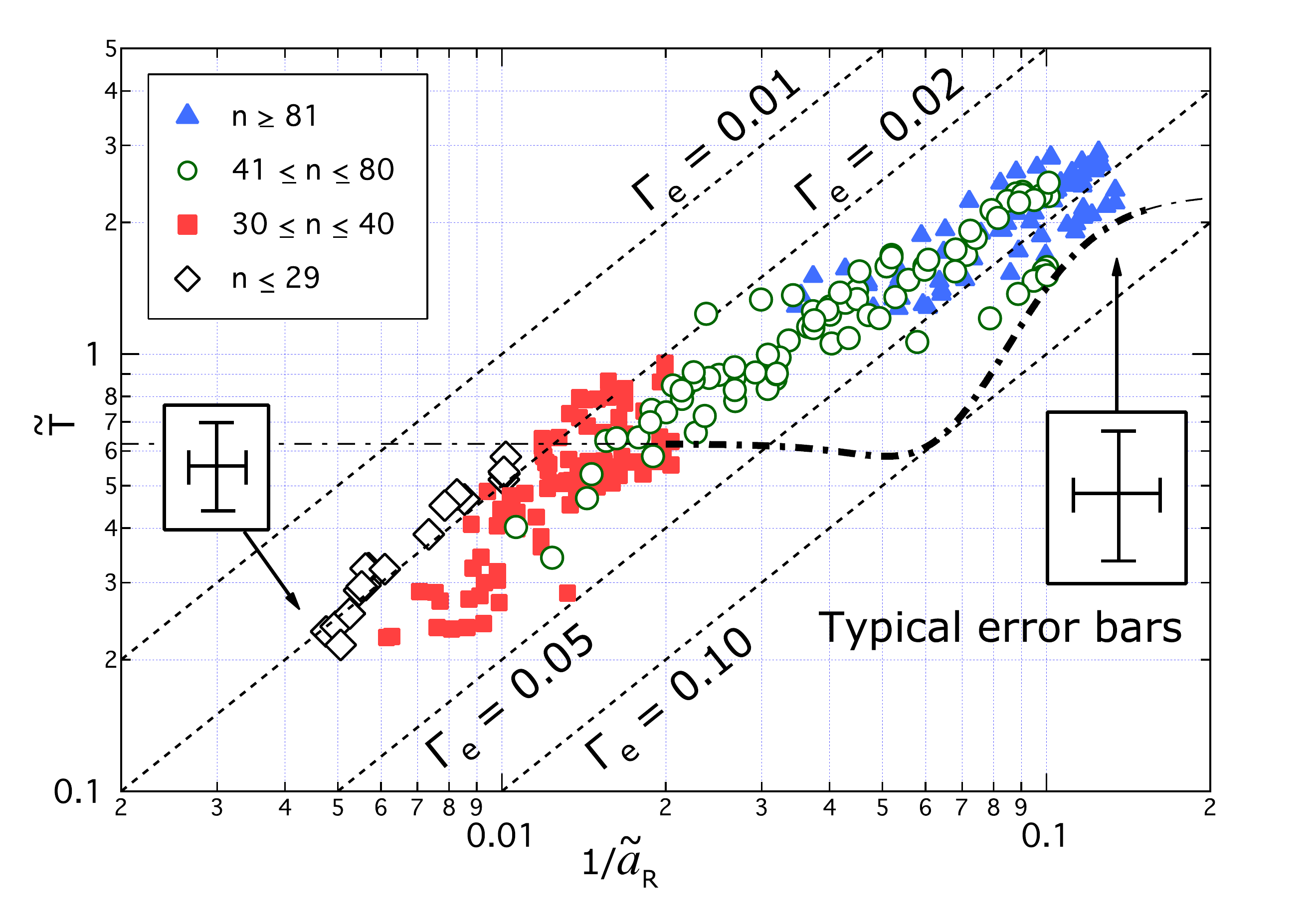}}}
\caption{Graph of experimental $\tilde T$ versus $1/\tilde a_R$ values for UNPs evolving from cold $nd$ Rb Rydberg samples in the range $24 \le n \le 120$. The data shown comprise 28 different $n$ values, and for each $n$, data were obtained for at least six, and up to 20, different densities. We distinguish the data in terms of ranges of $n$ as shown in the legend. (For reference, $n \le 29$ corresponds to $|E_{b,i}|/k_b > 200$ K; $30 \le n \le 40$ to $200 \ge |E_{b,i}|/k_b > 100$ K; $41 \le n \le 80$ to $100 \ge |E_{b,i}|/k_b > 25$ K; and $n \ge 81$ to $|E_{b,i}|/k_b \le 25$ K.)  Typical error bars are as shown. Also shown are black dashed lines corresponding to $\Gamma_e = 0.01, \ 0.02, \ 0.05$, and $0.1$, found assuming that $a_e = a_R$. The black $- \cdot -$ line is predicted using Eqs. \ref{fVs1OvaRheuristic} - \ref{ScaledT} as described in Section \ref{temp40} for $n > 40$ (the \textbf{bold} section corresponds to the range of final ionization fractions for which the model is valid).} 
\label{scaledTvsscaleda}
\end{figure}

For the moment, given the relative insensitivity of the ratio $a_e/a_R$ on $f$, we will find it useful to continue using $a_R$ as a proxy for $a_e$, and the related scaled quantity $\tilde a_R = a_R/2{n^\ast}^2a_0$ as a proxy for $\tilde a_e$. We have therefore plotted $\tilde T$ versus $1/\tilde a_R$, as shown in Fig. \ref{scaledTvsscaleda}. A full discussion of the analysis of this experimental data is described in Section \ref{anal}. First, however, there are a number of significant conclusions that can be drawn from the data shown in Fig. \ref{scaledTvsscaleda}. As can be seen, using the scaled quantities $\tilde a_R$ and $\tilde T$ results in a single universal curve, indicating that the system is behaving in an approximately classical fashion. This is a consequence of general scaling behaviors of processes involving Rydberg atoms, for instance ionization by collisions with electrons \cite{vrin05}, and the relative insignificance of deexcitation by radiative decay, at least for states with $n \ge 40$ \cite{bran10}. The values of $\tilde T$ are all of order unity ($0.2 \lesssim \tilde T \lesssim 3$), in line with what one would expect based on a consideration of the inverse process, TBR in UNPs made by photoionization, for which electrons with energy $k_BT_e$ undergo recombination into states bound by $\sim k_BT_e$ \cite{kill01, robx02}. Additionally, $\tilde T$ is larger for high $n$ states than for low $n$ states, and has an almost linear dependence on $1/\tilde a_R$. 

As with Fig.\,\ref{TeVsa}, in Fig.\,\ref{scaledTvsscaleda} the ratio of the Wigner-Seitz radius for the electrons in the UNP to the mean Rydberg atom separation varies with $f$ over the range of $\tilde a_R$ values. However, theoretical results given in Refs. \cite{robx02,robx03,poh03} show that a maximum of $f \approx 0.7$ of the Rydberg atoms ionize during the avalanche for principal quantum numbers in the range $n = 45 - 70$ and densities of $10^8 - 10^9$ cm$^{-3}$ (our own analysis, described in Section \ref{anal} gives a maximum final ionization fraction of $f_f = 0.83$ for $n = 120$). This suggests that, for the data with $1/\tilde a_R > 0.1$, $\tilde T > 2$ in Fig. \ref{scaledTvsscaleda}, $1/\tilde a_e = f^{1/3}/\tilde a_R \approx 0.9/\tilde a_R$. For these data points, the values of $\Gamma_e \approx 0.06$ found assuming $a_e = a_R$ are therefore quite accurate. On the other hand, in the low $1/\tilde a_R$, low $\tilde T$ part of the graph, the data points will be skewed to the left relative to those in Fig. \ref{scaledTvsscaleda} when plotted versus $1/\tilde a_e$. For instance, the lowest point on the graph in Fig. \ref{scaledTvsscaleda} is ($1/\tilde a_R \approx 0.005$, $\tilde T \approx 0.2$). This point would correspond to ($1/\tilde a_e \approx 0.001$, $\tilde T \approx 0.2$) if $f = 0.01$, giving $\Gamma_e \approx 0.005$.

The range of $\Gamma_e$ values we obtain are reasonably comparable with those reported in other experiments using different methods. Specifically, the NIST group found $0.02 < \Gamma_e < 0.08$ \cite{robe04} and $\Gamma_e = 0.13$ \cite{flet07} for Xe plasmas made by photoionization, using electron spilling, and measurement of TBR rates, respectively. Gupta \textit{et al.} found $\Gamma_e \lesssim 0.1$ for Sr plasmas made by photionization with $T_{e,0} > 45$ K, $\rho_{ion,avg} < 4.0 \times 10^9$ cm$^{-3}$ using the method of laser velocimetry of the Sr$^+$ resonance line, but found larger $\Gamma_e$ values at lower initial temperature and higher density \cite{gupt07}. Additionally, we have compared our $\Gamma_e$ values with those obtained from a Monte-Carlo model provided to us by Robicheaux \cite{robx02,robx03}. While the model analyzes UNPs made by photionization, rather than those which evolve from Rydberg samples, it is to be expected that there should be a reasonably smooth variation in the plasma properties in the region of the ionization limit. When we simulate this system using density and size parameters comparable to our experiment $\rho_{R,avg} \sim 10^8$ cm$^{-3}$ and $\sigma_0 \approx 400 \ \mu$m, we find $\Gamma_e$ values in the range from 0.01 ($T_{e,0} = 140$ K) to 0.09 ($T_{e,0} = 20$ K). On the other hand, UNPs which evolve from Rydberg states of cold NO molecules in a supersonic beam have been reported to have $T_e \approx 7$ K at a density such that $a_{e} = 360$ nm, implying $\Gamma_e \approx 7$ \cite{morr09}.

\section{{NUMERICAL MODELING}\label{anal}}
\subsection{{Approach and initial conditions}\label{scope}} 

In order to understand what the data shown in Fig. \ref{scaledTvsscaleda} say about how UNPs which evolve from cold Rydberg samples behave, and in particular, what determines the effective initial electron temperature, $T_{e,0}$, we have carried out extensive modeling of the interaction between a cold plasma and a co-existing reservoir of Rydberg atoms. Specifically, we have used a program provided to us by Robicheaux, which uses the Monte Carlo method to calculate the effects of electron-Rydberg collisions, TBR, and other interactions on the plasma evolution process \cite{robx02,robx03}. The initial conditions are specified numbers of ions, electrons with a specific temperature, and Rydberg atoms in a specific $nd$ state. The Rydberg atoms are distributed randomly within a Gaussian envelope with an initial characteristic radius $\sigma_0$. Similarly, the initial electron and ion density distributions are Gaussian with initial radius $\sigma_0$. For each electron-Rydberg collision, the probabilities for excitation, deexcitation, and ionization are compared with randomly generated probabilities, and the effect of the successful outcome is accounted for in terms of the change in the number of ions, free electrons, and neutral atoms, the energy of each atom, and the mean energy of the electrons. The effects of electron-ion collisions on the ion number, Rydberg atom number, their state distributions, positions, and velocities are tracked, as well as the effect of radiative decay of the Rydberg atoms. For each time iteration, the differential equations which describe the global plasma parameters (see Ref. \cite{robx03}, Eqs. 12) are solved numerically using the relevant particle numbers and energies. The program does not model how a cold Rydberg sample evolves into a UNP; rather, we use it to find how a reservoir of Rydberg atoms affects the evolution of a co-existing UNP. Hence, we are effectively modeling the evolution of a Rydberg plasma from the onset of the avalanche regime, and we run the simulation until a final time of 40 $\mu$s later. 

In understanding this approach to modeling of Rydberg plasmas, it is useful to make an analogy with how the electron temperature evolves in UNPs made by direct photoionization. Immediately after the ionization laser pulse, a plasma forms if there is sufficient ion density to trap the electrons. If the densities are high and the electron temperature $T_e$ is low, TBR, with a rate which scales as $\rho_e^2 \rho_{ion} T_e^{-9/2}$ ($\rho_e$ and $\rho_{ion}$ are the electron and ion density, respectively), heats the plasma electrons and creates bound Rydberg atoms. The TBR phase ends due to the $T_e^{-9/2}$ rate dependence: the remaining electrons are heated by TBR, which in turn reduces the TBR rate until it becomes comparable to the rate of electron replenishment due to ionizing electron-Rydberg collisions \cite{flet07}. Thereafter, the plasma expands, and the electrons cool adiabatically, though a small amount of electron heating occurs due to electron-Rydberg collisions driving the atoms to more deeply bound states, and TBR, which continues at a low rate because $T_e$ falls throughout the expansion \cite{flet07}. The asymptotic plasma expansion velocity, $v_0$, is described by Eq. \ref{vZero}, where $T_{e,0}$ is the electron temperature resulting from the ionizing photon's excess energy, heat added during the TBR phase, and the small amount of heating or cooling which happens after the TBR phase ends \cite{kul00,kill01}. 

For Rydberg plasmas, the period which corresponds to the TBR phase is the avalanche regime, where the rates of collisions between electrons and Rydberg atoms, and electrons and ions, are high. During this phase, for $n \ge 40$, anywhere from a few percent to more than 80\% of the atoms ionize, while the remaining bound atoms are scattered to more deeply bound states. However, there will be significant interaction between the electrons and Rydberg atoms throughout the evolution of a Rydberg plasma, given that there is a much larger reservoir of Rydberg atoms than in a photoionization-initiated UNP \cite{poh03}. The avalanche regime is thus unlikely to have as distinct an end point as the TBR phase in conventional UNPs. As noted above in Section \ref{expt}, our simulations showed that $v_0$ continued to increase for several tens of $\mu$s into the plasma evolution. Collisions between electrons and the Rydberg atoms drive the atoms to more deeply bound states, and the energy so liberated accelerates the plasma expansion \cite{robx02}. We chose an end point for the simulations of 40 $\mu$s after the plasma was created as a reasonable compromise for comparison with the experimental results. At this time we found that the quantity $T_{e,0} = m_{ion}v_0^2/k_B$ attained at least 95\% of the value it would have reached for much longer simulation times. In a conventional UNP made by photoionization, and which experiences no TBR heating, $T_{e,0}$ would be the actual initial electron temperature; however, in our case, it is just a useful measure of the net thermal energy transferred from other degrees of freedom into the outward expansion of the plasma.


For the moment, we will concentrate on plasmas which evolve from Rydberg atoms with $n \gtrsim 40$. (Below $n = 40$, we find that there is in general very little additional ionization after the threshold condition is reached. This regime is discussed more extensively in Section \ref{templow}.) With the picture described in the previous paragraph in mind, we consider the initial condition to be the beginning of the avalanche regime. At this time, when the initial ionization fraction is $f_i$, there are $N_{R,i} = (1 - f_i) N$ Rydberg atoms in a specific $nd$ state with binding energy $E_{b,i}$ which interact with $N_{ion,i} = f_{i} N$ ions and $N_{e,i} = N_{ion,i}$ electrons with a specified initial temperature, $T_{e,i}$. (The ions and Rydberg atoms are assumed to be stationary at this time.) The final condition is the end of the simulation, 40 $\mu$s after the plasma is created. At this time, the ionization fraction is $f_f = N_{ion,f}/N$, the mean Rydberg binding energy is $\bar E_{b,f}$, and the electron temperature is $T_{e,f}$. Additionally, thermal and binding energy have been converted to kinetic energy of the ions. When the plasma expansion velocity is $v_0$, the mean ion kinetic energy is $(3/2) \, m_{ion} \, v_0^2$ (see Section \ref{temp40}). We obtain the plasma expansion velocity for each time step in the evolution using the characteristic radius of the UNP, $\sigma(t)$, which is one of the program outputs, and the equation $\sigma(t) = \sqrt{\sigma^2_0 + v^2_0 \, t^2}$. The value of $v_0$ for the last time iteration before the simulations end at 40 $\mu$s is the one we relate back to the effective initial electron temperature, $T_{e,0}$, using Eq. \ref{vZero}, again assuming that $T_{ion,0} = 0$. 

Our assumptions are crude in that they do not consider that the initial ionization process which seeds the plasma probably leads to a distribution of electron energies and redistributes some of the Rydberg atoms to states different from that excited by the laser \cite{robx05,bet09a}. Additionally, cold dipole collisions are faster for close atom pairs causing a Rydberg atom distribution different from one which is random within a Gaussian envelope \cite{sade14}. However, these affect only a fraction $\sim f_i$ of the Rydberg atoms, and the net effect of neglecting these deviations at the end of the simulation is probably negligible. While we were unable to test this hypothesis with regard to the reduction in the number of close atom pairs, the simulation results were not significantly dependent on the value of $T_{e,i}$ used, and varied only weakly with $E_{b,i}$.

\begin{figure*}
\centerline{\resizebox{1.00\textwidth}{!}{\includegraphics{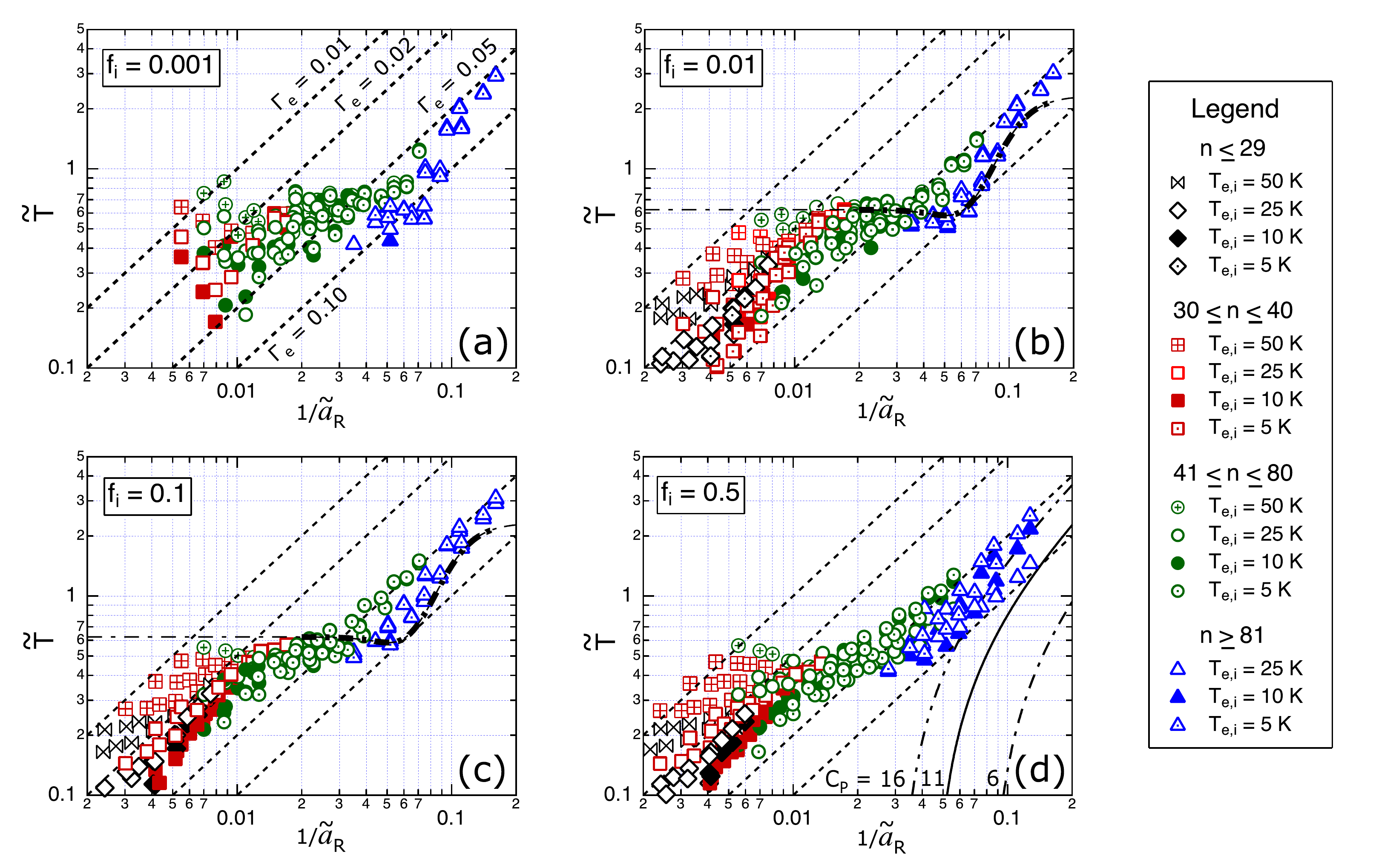}}}
\caption{Results of numerical modeling for $\tilde T$ versus $1/\tilde a_R$ for UNPs evolving from cold $nd$ Rb Rydberg samples in the range $24 \le n \le 120$. (a) $f_i = 10^{-3}$; (b) $f_i = 10^{-2}$; (c) $f_i = 10^{-1}$; and (d) $f_i = 0.5$. We distinguish the data in terms of different $n$ (initial binding energy) ranges and $T_{e,i}$ values as shown in the legend. (For reference, $n \le 29$ corresponds to $|E_{b,i}|/k_b > 200$ K; $30 \le n \le 40$ to $200 \ge |E_{b,i}|/k_b > 100$ K; $41 \le n \le 80$ to $100 \ge |E_{b,i}|/k_b > 25$ K; and $n \ge 81$ to $|E_{b,i}|/k_b \le 25$ K.) Where the 10 K and 25 K data points appear to be missing, they lie underneath the corresponding 5 K symbol. The axes have the same range as those of the graph shown in Fig. \ref{scaledTvsscaleda}. The black $- \cdot -$ line in (b) and (c) is predicted using Eqs. \ref{fVs1OvaRheuristic} - \ref{ScaledT} as described in Section \ref{temp40} for $n > 40$ (the \textbf{bold} section corresponds to the range of final ionization fractions for which the model is valid, $1.5 \times f_i \le f_f \le 0.83$). Also shown are lines corresponding to $\Gamma_e = 0.01, \ 0.02, \ 0.05$, and $0.1$, found assuming that $a_e = a_R$, with the values of $\Gamma_e$ for each shown in (a). In addition, (d) shows parameter space limits due to threshold lowering for $C_P = 6$ (black, $-$ - $-$), $C_P = 11$ (black solid line) and $C_P = 16$ (black, $-$ - - $-$) - see Section \ref{temp40} for details.}
\label{thScaledTvsscaleda}
\end{figure*}

While the quantities $f_i$ and $T_{e,i}$ are presumably determined by the dominant initial ionization mechanism, they are not quantities which we can measure and they are not predicted by the program we use. Instead, for a given $nd$ state, we have run the models with $f_i = 0.5, \ 0.1, \ 0.01, \ 0.001$, and $T_{e,i} = 5$ K ($n \ge 50$ only), 10 K, 25 K (all $n$), and 50 K ($n \le 40$ only). It turned out that the results for almost all the Rydberg states we investigated were substantially independent of the $f_i$ and $T_{e,i}$ values chosen. This makes sense: given the large number of Rydberg atoms relative to the number of seed electrons, after only a few electron-atom collisions, the properties of the free electrons are determined by the Rydberg reservoir much more than by the initial electron temperature and density. In addition, all the simulation results for the relationship between $\tilde T$ and $1/\tilde a_R$ closely agreed with what we measured, as can be seen in Fig. \ref{thScaledTvsscaleda}. The $f_i$ and $T_{e,i}$ values we used in our simulations are reasonable based the initial ionization mechanisms \cite{robx03,poh03,li04,vit82,robx05,spen82b,bet09c} and the number of ions needed to establish a potential well of depth $\sim |E_{b,i}|$ and radius $\sigma_0$ \cite{kill99}. We will therefore consider the insensitivity of the simulation results to specific $f_i$ and $T_{e,i}$ values and their agreement with the experimental results as a sufficient justification for our choice of initial parameters.

For each $nd$, $f_{i}$, $T_{e,i}$ combination, we use average total densities (atoms + ions) $N/(4 \pi \sigma_0^2)^{3/2} = 1 \times 10^7, \, 2 \times 10^7, \, 3 \times 10^7, \, 5 \times 10^7, \, 1 \times 10^8, \, 2 \times 10^8$, and $3 \times 10^8$ cm$^{-3}$, and an initial plasma radius $\sigma_0 = 212$ $\mu$m. (The maximum density and the $\sigma_0$ value we use are determined by available computing power.) The plasmas evolve for 40 $\mu$s, at which time we evaluate $T_{e,0}$ and $f_f$, as well as $|\bar E_{b,f}|$. This latter quantity is found by averaging the the energies of all the neutral atoms with $n \ge 5$; however, the fraction of atoms which end up below $n=5$ is significant only for initial $nd$ states with $n \le 35$. (States with $n < 5$ are would be inaccessible for Rb if the calculation was fully quantum mechanical. However, the model is semi-classical, and the $n < 5$ criterion matters only in that these atoms are removed from the system being modeled \cite{robx03}. Practically, even atoms significantly above $n = 5$ have very little impact on the plasma dynamics, given their high radiative decay rates and small geometric cross sections.) 

\subsection{{Results of modeling: $\tilde T$ versus $1/\tilde a_R$}\label{thres}}

The results of the analysis described above for the behavior of $\tilde T$ versus $1/\tilde a_R$ are shown in Fig. \ref{thScaledTvsscaleda}. As can be seen, for one particular $f_i$ value, there is substantial agreement of the results for different $T_{e,i}$ values for $n > 40$. On the other hand, for $n \lesssim 40$, higher $T_{e,i}$ (50 K) values result in significantly higher $\tilde T$ values than for $T_{e,i} = 5$ K, 10 K and 25 K, which give consistent $\tilde T$ values. This regime is discussed in Section \ref{templow}. 

In comparing the graphs with different initial ionization fractions, it can be seen that the relationship between  $\tilde T$ and $1/\tilde a_R$ is substantially independent of the value of $f_i$. The agreement of the simulations using different $f_i$ is strongest for $n > 40$, but even for $n \lesssim 40$ and $T_{e,i} = 10$ K and 25 K, there is good agreement of the results. For $f_i = 10^{-3}$, we were not able to obtain as many results as for the other $f_i$ values. Many of these simulations failed due to insufficient electron and ion densities to sustain a plasma. Given the instability of the $f_i = 10^{-3}$ simulations, and the low likelihood that a situation where $f_i = 0.5$ would arise, much of the discussion below concentrates on $f_i = 10^{-1}$ and $f_i = 10^{-2}$.

Comparing the experimental data in Fig. \ref{scaledTvsscaleda} with the results of the modeling shown in Fig. \ref{thScaledTvsscaleda}, there is strong qualitative and quantitative agreement. The endpoints of the range of ($1/\tilde a_R$, $\tilde T$) coordinates, anchored at the low $\tilde T$ end by the $24d$ data and on the high $\tilde T$ end by $120d$, match reasonably well, and the range of $\Gamma_e$ values (using the proxy $\tilde a_e = \tilde a_R$) is similar. On the other hand, there are significant differences between the predictions of the model, and the experimental results. First, the $f_i = 10^{-1}$ and $10^{-2}$ simulations exhibit a weak plateau where $\tilde T \approx 0.6$ for $0.01 \lesssim 1/\tilde a_R \lesssim 0.03$, whereas in the experimental data, there is no evidence of such a feature. (On the other hand, the plateau behavior is limited to states with $n \ge 80$, and the simulations for $n < 80$ are actually very similar to the experimental results.) Additionally, the range of ($1/\tilde a_R$, $\tilde T$) values exhibited by a given $n$ state are generally higher for the experimental data than in the simulations. This discrepancy is primarily caused by the difference in the densities used in the simulations from what was achieved in the experiment. Consequently, the upper and lower bounds of the range of $1/a_R$ values were approximately 50\% higher in the simulations than in the experiment. Additionally, it is likely that the difference in $\sigma_0$ values (212 $\mu$m in the simulations, 400 $\mu$m in the experiment) contributes to this difference: in our simulations of UNPs created by direct photoionization, we found that $v_0$ increased with smaller $\sigma_0$ values for low $T_{e,0}$ where TBR is significant (the other conditions were kept unchanged).


\subsection{{Evolution behavior of a Rydberg plasma}\label{mech}}

\begin{figure}
\centerline{\resizebox{0.5\textwidth}{!}{\includegraphics{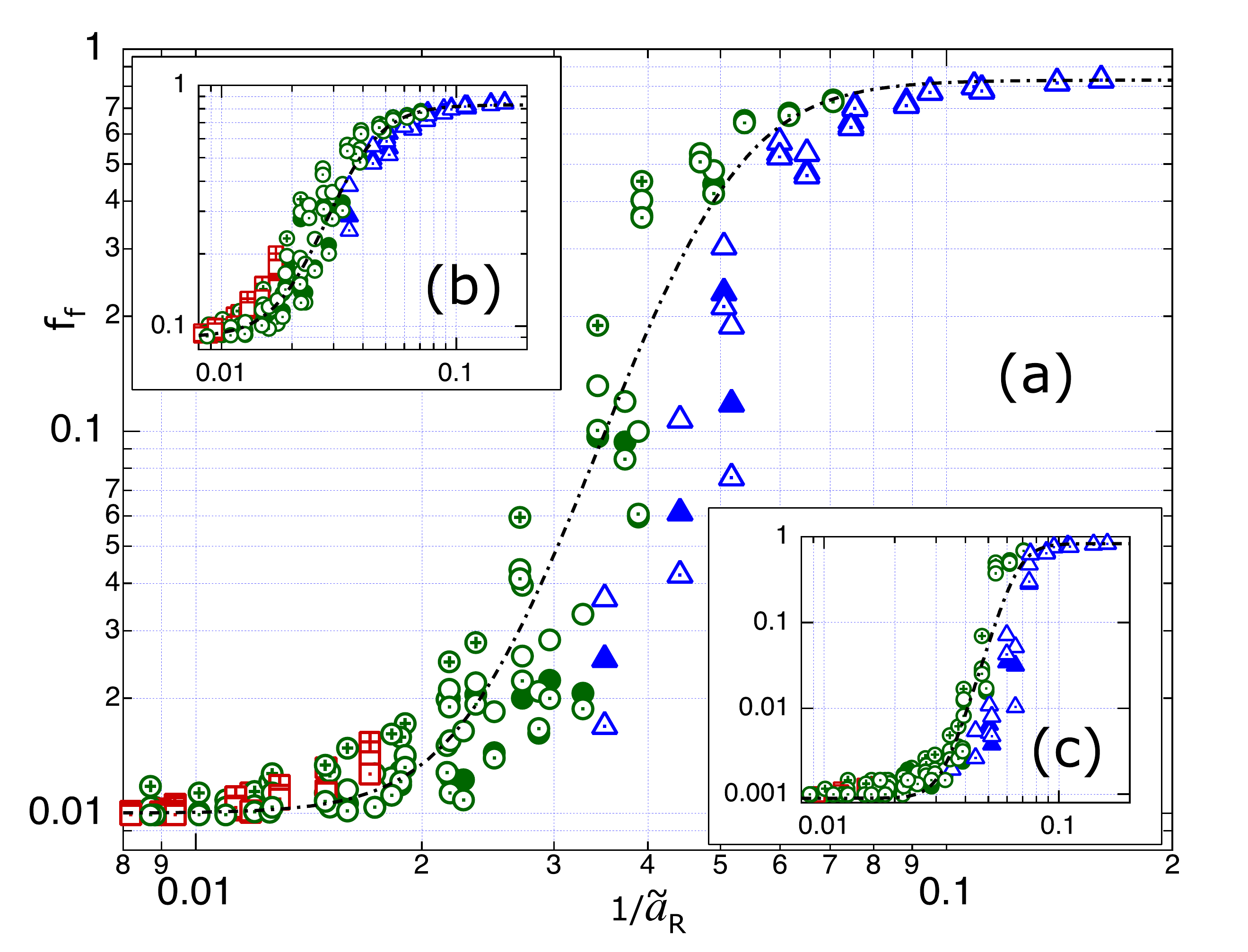}}}
\caption{Results of numerical modeling for $f_f$ versus $1/\tilde a_R$ for UNPs evolving from $nd$ cold Rb Rydberg samples in the range $24 \le n \le 120$. (a) $f_i = 10^{-2}$; insets (b) $f_i = 10^{-1}$ and (c) $f_i = 10^{-3}$. We distinguish the data in terms of ranges of $n$ and $T_{e,i}$ using the same scheme as in Fig. \ref{thScaledTvsscaleda}. The black dashed lines are a simple heuristic relationship given by Eq. \ref{fVs1OvaRheuristic}.}
\label{thFfVs1OvScaleda}
\end{figure}

The similarity of the curves shown in Figs. \ref{thScaledTvsscaleda}(a), (b), and (c), and to a lesser extent, (d), for vastly different $f_i$ and for all initial $T_{e,i}$ in the region $1/\tilde a_R > 0.02$ ($n \gtrsim 40$), and their similarity to the experimental data shown in Fig. \ref{scaledTvsscaleda}, is a significant result. Basically, it suggests two conclusions. The first is that such Rydberg plasmas, once they reach threshold, evolve in a manner which is independent of the initial plasma seeding mechanism. As noted above, this is because the electron temperature and density, and the Rydberg state distribution all change rapidly due to the high electron-Rydberg collision rate at the onset of the avalanche.
The second conclusion suggested by Figs. \ref{thScaledTvsscaleda}(a), (b), and (c) is that for $1/\tilde a_R > 0.02$, $T_{e,0}$ for the UNP must be intimately related to the final ionization fraction, $f_f$. To test this hypothesis, we have looked at how $f_f$ depends on $1/\tilde a_R$ for $f_i = 0.1, \, 0.01$, and 0.001. These results are shown in Fig. \ref{thFfVs1OvScaleda}. As can be seen, the final ionization fraction, $f_f$, remains smaller than twice the initial value unless $1/\tilde a_R \gtrsim 0.02$, and this onset seems to correlate with the rise in $\tilde T$ seen at the same $1/\tilde a_R$ in Fig. \ref{thScaledTvsscaleda}. However, the ionization fraction then rises rapidly, reaching $f_f = 0.50$ between $1/\tilde a_R = 0.04$ and 0.07, and saturates at $f_f \approx 0.83$ regardless of $f_i$. There is significant variation in $f_f$ values for different $T_{e,i}$ and different $|E_{b,i}|$ in the transition region; however, regardless of these two parameters, the transition from low to high $f_f$ occurs in a well-defined range of $1/\tilde a_R$ values, and the range varies only slightly with $f_i$. Additionally, the region where $f_f/f_i \ge 2$ is exhibited only for initial states with $n > 40$, and for all initial states with $n \le 40$, the number of Rydberg atoms which ionize during the avalanche is very low. 

We have found a simple heuristic relationship between $f_f$ and $1/\tilde a_R$ which describes the gross features of the variation of $f_f$ with $1/\tilde a_R$:
\begin{equation}
f_f = f_i + (0.83 - f_i) \times \frac{k/\tilde a_R^m}{1 + k/\tilde a_R^m}. \label{fVs1OvaRheuristic}
\end{equation}
Values of $k$ and $m$ which give a reasonable description to the data shown in Fig. \ref{thFfVs1OvScaleda} are: $f_i = 0.1$, $k = 5.4 \times 10^5$ and $m = 4$; $f_i = 0.01$, $k = 6.6 \times 10^7$ and $m = 6$; and $f_i = 0.001$, $k = 3.6 \times 10^{10}$ and $m = 9$. These curves are shown in Fig. \ref{thFfVs1OvScaleda}. As can be seen, Eq. \ref{fVs1OvaRheuristic} does not describe well the $T_{e,i}$- and $|E_{b,i}|$-dependent variations in the transition region, nor does it work well when $f_f \approx f_i$. However, it is a useful relation which we will use in Section \ref{temp40}.

\begin{figure}
\centerline{\resizebox{0.5\textwidth}{!}{\includegraphics{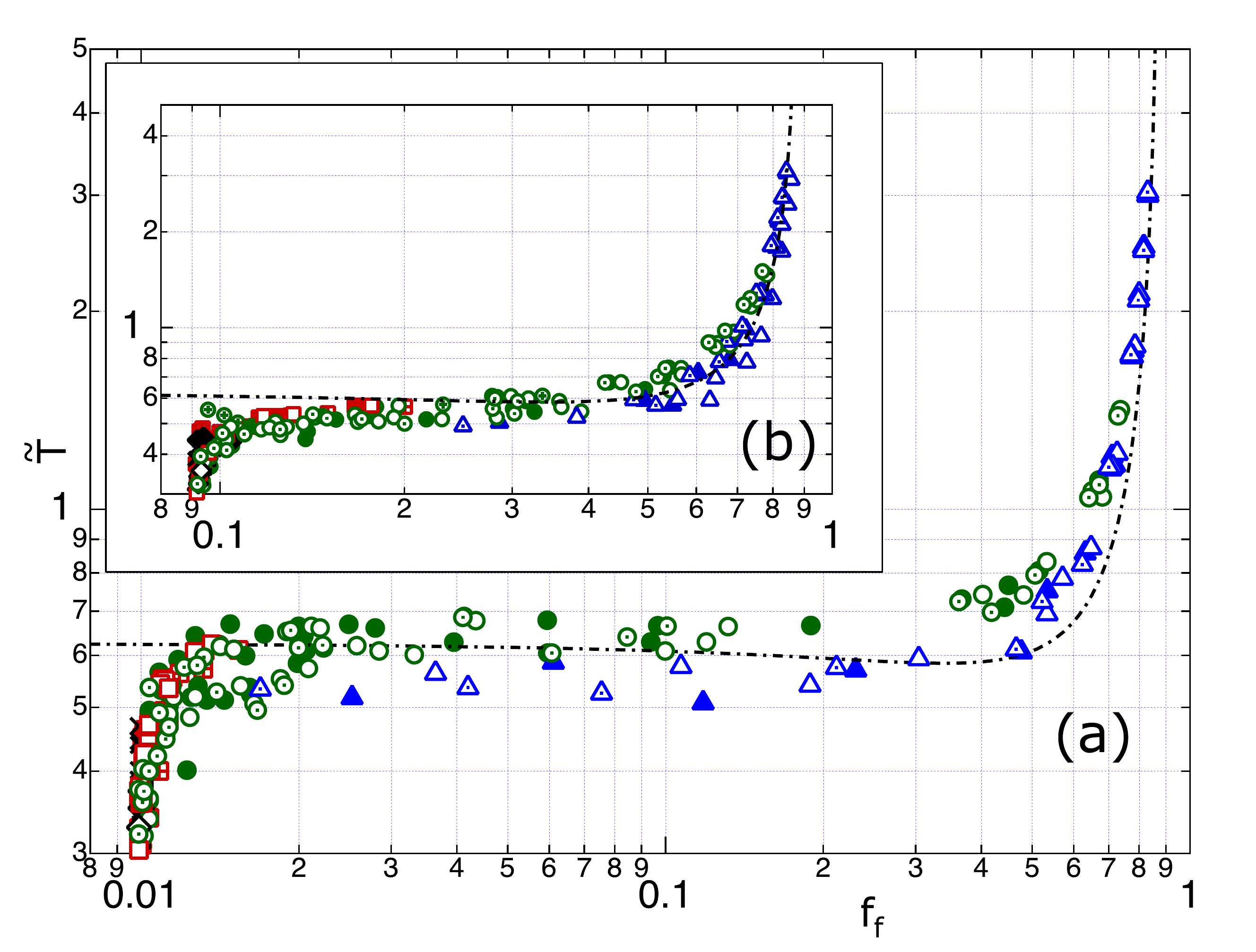}}}
\caption{(a) Results of numerical modeling for $\tilde T$ versus $f_f$, for $f_i = 0.01$; inset (b), for $f_i = 0.1$. We distinguish the data in terms of ranges of $n$ and $T_{e,i}$ using the same scheme as in Fig. \ref{thScaledTvsscaleda}. The black $- \cdot -$ lines are obtained using Eqs. \ref{betaheur} and \ref{ScaledT} as described in Section \ref{temp40} and Eq. \ref{ScaledT}.}
\label{thScTVsFf}
\end{figure}

\begin{figure}
\centerline{\resizebox{0.5\textwidth}{!}{\includegraphics{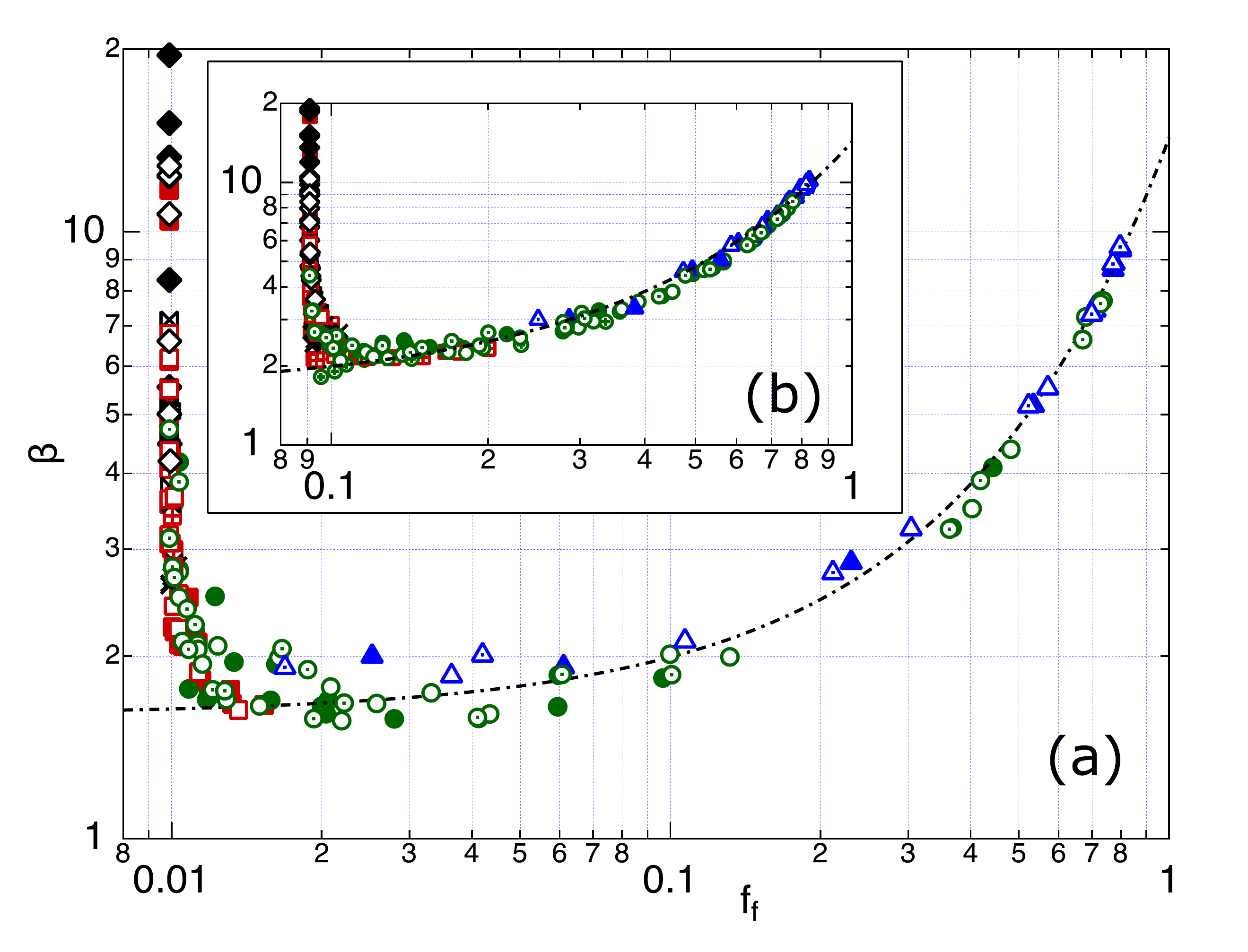}}}
\caption{(a) Results of numerical modeling for $\beta$ versus $f_f$, for $f_i = 0.01$; inset (b), for $f_i = 0.1$. We distinguish the data in terms of anges of $n$ and $T_{e,i}$ using the same scheme as in Fig. \ref{thScaledTvsscaleda}. The black $- \cdot -$ lines are the heuristic $\beta(f_f)$ given by Eq. \ref{betaheur}.}
\label{thBetaVsFf}
\end{figure}

Comparing the results shown in Figs. \ref{thScaledTvsscaleda} and \ref{thFfVs1OvScaleda} in the region $1/\tilde a_R > 0.02$, we see that $f_f$ rises rapidly with increasing $1/\tilde a_R$, and $\tilde T$ shows a marked increase also. There is significant scatter in the dependence of both $f_f$ and $\tilde T$ on $1/\tilde a_R$ for $0.02 \le 1/\tilde a_R \le 0.1$, but the scatter is markedly less for both $f_f$ and $\tilde T$ in the region $1/\tilde a_R > 0.1$. This suggests that there is a strong correlation between $\tilde T$ and $f_f$, and the scatter of $\tilde T$ versus $1/\tilde a_R$ in Fig. \ref{thScaledTvsscaleda} is related to the scatter of $f_f$ versus $1/\tilde a_R$ in Fig. \ref{thFfVs1OvScaleda}. To test this relationship, we looked at the dependencies of $\tilde T$ and the parameter $\beta \equiv |\bar E_{b,f}|/k_B T_{e,0}$ on $f_f$, where $|\bar E_{b,f}|$ is the average value of the binding energies of all atoms with $n > 5$ after 40 $\mu$s of plasma evolution time. These graphs are shown in Figs. \ref{thScTVsFf} and \ref{thBetaVsFf}. As can be seen, the hypothesis that there are well-defined relationships between $\tilde T$ and $f_f$, and between $\beta$ and $f_f$, is correct. This makes sense, since $\tilde T$, $\beta$, and $f_f$ are all determined by electron-Rydberg collisions. The greater the number of collisions, the greater the degree of ionization, and the more energy is transferred from the Rydberg atoms to the plasma expansion. 

We have used the data in Fig. \ref{thBetaVsFf} to find a simple heuristic relationship between $\beta$ and $f_f$. That relationship, which works well for both $f_i = 0.1$ and 0.01 provided that $f_f \gtrsim 2 \times f_i$, is 
\begin{equation}
\beta = 1.60 \, e^{2.19f_f}. \label{betaheur}
\end{equation}
Equation \ref{betaheur} is plotted along with the data in Fig. \ref{thBetaVsFf} (the black $- \cdot -$ lines). For low $f_f$ ($f_i < f_f \lesssim 0.3$), $\beta$ is in the range $1.6 < \beta < 3$, but as $f_f$ rises to its maximum value $\beta \rightarrow 10$. The high $\beta$ limit is reached only for $n \ge 100$, and corresponds to the maximum possible $f_f \approx 0.83$. At this point, for every six Rydberg atoms in the initial sample, five have ionized. To maintain the energy balance, the one remaining neutral atom must be much more deeply bound than the initial state, $|\bar E_{b,f}| \ge 6 \times |E_{b,i}|$. This limit is analogous to what has been found for low $T_{e,0}$ UNPs made by photoionization, where the Rydberg energy states formed by TBR are more deeply bound than for higher $T_{e,0}$ plasmas. Low $T_{e,0}$ photoionization-initiated UNPs and UNPs which evolve from Rydberg states with low $|E_{b,i}|$ expand slowly, and thus there is more time for electron-Rydberg collisions, which primarily lead to deexcitation of the atoms, to occur \cite{kill01,robx03}.

\subsection{{The effective initial electron temperature of an ultra cold Rydberg plasma when $n > 40$}\label{temp40}}
We show here that the connection between $\tilde T$ and $\beta$ is a consequence of energy conservation in the plasma evolution. It has been shown in numerous theoretical studies that there exists a bottleneck energy, $E_{bn}$, of the Rydberg state distribution for atoms in equilibrium with a plasma with electron temperature $T_e$ \cite{mans69,kuz02b,robx03,poh08,bann11}. Specifically, $E_{bn} \approx 4k_BT_e$, and Rydberg states with binding energy $|E_b| < E_{bn}$ will eventually ionize due to electron collisions, while those with $|E_b| > E_{bn}$ will be deexcited to states with lower $n$ which will eventually decay radiatively to the ground state. The seed electrons which result from BBR photoionization \cite{spen82b,bet09c}, cold dipole-dipole collisions \cite{robx05}, or by hot-cold Rydberg collisions \cite{vit82} all have distributions such that a significant fraction will have energies that are greater than $|E_{b,i}|/4$. This corresponds to the situation where the initial electron temperature $T_{e,i} > |E_{b,i}|/4 \, k_B$, and thus the electron temperature is more than sufficient at the beginning of the avalanche to ionize the Rydberg atoms in the original state, as well as many of the partner atoms deexcited by cold dipole collisions \cite{robx05}, because $E_{b,i}$ is above the bottleneck energy characteristic of a plasma with electron temperature $T_{e,i}$. As the plasma evolves, the interplay between electron-Rydberg exciting, deexciting, and ionizing collisions, and recombination maintains the energy balance in the evolution so that the magnitude of the average binding energy of the un-ionized atoms increases. The energy so liberated drives the plasma expansion so that the UNP achieves a final expansion velocity $v_0$. 

We can verify this picture by looking at what energy conservation predicts about the relationship between $\tilde T$, $\beta$, and $f_f$. Specifically, if we ignore the energy added to the system by BBR and hot-cold Rydberg collisions, the initial energy of the system is 
\begin{equation}
E_i = -(1-f_i) N |E_{b,i}| + f_i N \frac{3}{2} k_B  T_{e,i}, \label{Ei}
\end{equation}
where $N$ is the total number of atoms and ions. (We neglect the thermal energy of the ions; as discussed in the Introduction, $T_{ion,0} \lesssim 1$ K, and the subsequent adiabatic expansion causes the ion temperature to decrease further.) The final energy of the system is 
\begin{equation}
E_f = -(1-f_f) N |\bar E_{b,f}| + f_f N \frac{3}{2} k_B T_{e,f} + f_f N \frac{3}{2} m_{ion} \gamma^2 \sigma^2 \label{Ef}
\end{equation}
(see Ref. \cite{kill07}, Eq. 31e). The last term is the kinetic energy contained in the radial expansion of the ions (this is obtained by averaging the quantity $(1/2) \, m_{ion} \, |\, \vec u \, |^2$, where $\vec u$ and $\gamma (t)$ are given by Eq. \ref{ionvel}, over the density distribution of the ions \cite{robx03,kill07}). For times late in the plasma evolution, we can neglect $T_{e,f}$ because of adiabatic cooling. (In the simulations, $T_{e,f}$ was usually $\le 2$ K, and the maximum value we observed was 4 K.) We also make the approximation $m_{ion} \, \gamma^2 \, \sigma^2 \approx m_{ion} \, v_0^2 = k_B \, T_{e,0}$ which is valid for $v_0 \, t \gg \sigma_0$ (see Eq. \ref{ionvel}).

Equating the initial and final energies, and using $|\bar E_{b,f}| = \beta k_BT_{e,0}$, we obtain the following relationship
\begin{equation}
\tilde T = \frac{k_BT_{e,0}}{|E_{b,i}|} = \frac{( 1 - (1 + \frac{3}{2} \, \frac {k_B T_{e,i}}{|E_{b,i}|}) \, f_i)}{((1-f_f)\beta - \frac{3}{2}f_f )}. \label{ScaledT}
\end{equation}
For $f_i = 0.1$ the term in the numerator differs significantly from unity, and lies in the range $0.57$ (for $T_{e,i} = 25$ K and $n = 120$) to $0.90$ (the low $n$ limit). For $f_i = 0.01$, it lies in the range 0.96 - 0.99. We will ignore this dependence of $\tilde T$ on $T_{e,i}$ and $|E_{b,i}|$ and set the numerator to unity, which for $f_i = 0.1$ means that the $\tilde T$ values obtained are significant overestimates at high $n$, though for low $n$ they are good to within 20\%. For $f_i = 0.01$, the effect of setting the numerator equal to unity is negligible. 

We can now test the thesis that $\tilde T$ and $\beta$ are related by energy conservation. Specifically, we have substituted Eq. \ref{betaheur} into Eq. \ref{ScaledT} (with the numerator equal to unity), and this curve is plotted with the numerical data in Fig. \ref{thScTVsFf} (the black $- \cdot -$ lines). As can be seen, the agreement of the numerical data with the prediction based on Eqs. \ref{betaheur} and \ref{ScaledT} is very good. We can also understand why there are systematic differences between the heuristic and the numerical results. First, the effect of approximating the numerator to unity in Eq. \ref{ScaledT} can clearly be seen in Fig. \ref{thScTVsFf}(b) for $f_i = 0.1$: almost all the simulation data points lie below the heuristic, by $\lesssim 20$\% for $f_f \le 0.4$, but by significantly more for $f_f \ge 0.6$, as expected. On the other hand, this difference is much less apparent for $f_i = 0.01$ in Fig. \ref{thScTVsFf}(a). The second difference is that for both $f_i = 0.01$ and $0.1$, many data points have $\tilde T$ values which are greater than the heuristic predicts in the region $f_f \ge 0.4$. This is a manifestation of the fact that Eq. \ref{ScaledT} does not consider the kinetic energy of the Rydberg atoms. Since the atoms are assumed to be initially stationary in the simulations, the only atoms with significant kinetic energy at 40 $\mu$s will be the ones formed by recombination of electrons with ions which have already acquired a significant outward velocity. In Ref. \cite{poh03}, it is argued that during the expansion, a given Rydberg atom will ionize and recombine many times during the evolution of a Rydberg UNP. Consequently, all of the atoms are essentially dragged along by the plasma, and there should therefore be an additional term in the final energy that is $\lesssim (1 - f_f )\, N \, (3/2) \, m_{ion} \, v_0^2$, equivalent to $\lesssim (1 - f_f )\, N \, (3/2) \, k_B \, T_{e,0}$. This affects the $-(3/2) \, f_f$ term in the denominator of Eq. \ref{ScaledT}, and if the Rydberg atoms have the same final velocity distribution as the ions, this term becomes $-(3/2)$, leading to $\tilde T$ values which are higher than Eq. \ref{ScaledT} predicts. There is no simple relationship we can use to estimate the amount of kinetic energy the Rydberg atoms acquire during the expansion, other than the upper limit, which would correspond to $\tilde T \approx 5$ at $f_f = 0.83$ (i.e., replacing $-(3/2) \, f_f$ with $-(3/2)$ in Eq. \ref{ScaledT}). However, the effect will only influence $\tilde T$ values where $f_f$ is significant ($f_f \gtrsim 0.1$, say) - if there isn't much ionization, there can be few recombination events that result in moving Rydberg atoms. Accounting for the Rydberg atom kinetic energy, the $\tilde T$ values could be as much as twice those estimated using Eq. \ref{ScaledT} in this region. 

We point out in passing that our $\beta$ and $\tilde T$ values are consistent with the theoretical analysis reported in Ref. \cite{poh03}. While this paper reports only values for ionization fractions and Rydberg state distributions as functions of time when an $n = 70$ state is initially excited, and does not give the corresponding information about the electron temperature, there is enough information in Fig. 3 in Ref. \cite{poh03} to extract a value for $\beta \tilde T$. Specifically, at an evolution time of 25 $\mu$s Pohl \textit{et al.} found $|\bar E_{b,f}|/|E_{b,i}| = \beta k_B T_{e,0}/|E_{b,i}| = \beta \tilde T \approx 8$ when $f_f \approx 0.7$ and $\rho_{avg} \approx 5 \times 10^9$ cm$^{-3}$. While this density is significantly higher than the one used in our analysis, using Figs. \ref{thScTVsFf} and \ref{thBetaVsFf}, we obtain $\beta \approx 7.5$ and $\tilde T \approx 0.9$, giving a value for $\beta \tilde T \approx 7$ when $f_f = 0.7$. In addition, there is agreement with the experiment reported in Ref. \cite{rdsv13}, where they found $\tilde T = 0.6 \pm 0.2$, though those reported in Ref. \cite{sier14}, $\Tilde T = 0.29$ and 0.46 for plasmas from $45s_{1/2}$ and $40d$ respectively, are somewhat lower than the minimum value of 0.6 that we get from Eq. \ref{ScaledT} using the minimum $\beta$ value consistent with Fig. \ref{thBetaVsFf}. On the other hand, the work reported in Ref. \cite{rdsv13} claims a value of $f_f \sim 1$, and for Ref. \cite{sier14}, it was likely similar. For such high $f_f$, our analysis suggests $\tilde T \approx 2.5$. However, both experiments used optical dipole traps which give small $\sigma_0$ and make high densities possible, leading to long-lived plasmas ($\lesssim 400 \ \mu$s in \cite{sier14}). Additionally, both experiments excited the Rb $5p_{3/2} \rightarrow n\ell$ transitions for much longer time durations (several $\mu$s to 200 $\mu$s) than we do ($\sim 5$ ns). The fact that the Rydberg state population changes during the plasma evolution due to this replenishment, and not just as a consequence of the interaction between the plasma and the Rydberg atoms, likely makes the dynamical behaviors seen in \cite{rdsv13,sier14} significantly different to ours. 

We have also used the heuristic relationship between $f_f$ and $1/\tilde a_R$ described by Eq. \ref{fVs1OvaRheuristic}, and Eq. \ref{ScaledT} to model the dependence of $\tilde T$ on $1/\tilde a_R$. Specifically, we used the constants for the condition $f_i = 0.01$ in Eq. \ref{fVs1OvaRheuristic}, and this curve is shown in Fig. \ref{scaledTvsscaleda} and Figs. \ref{thScaledTvsscaleda}(b) and (c). (The part of the curve denoted by the bold line in the figures corresponds to final ionization fractions for which the heuristics are reasonably accurate representations of the numerical results, i.e., the range $2 \times f_i \le f_f \le 0.83$.) The differences in the $f_f$ versus $1/\tilde a_R$ curves for $f_i = 0.1$ and $0.01$ affect only how $\tilde T$ changes with $1/\tilde a_R$ in the region above $1/\tilde a_R = 0.04$. 

In the experimental data shown in Fig. \ref{scaledTvsscaleda}, we see that $\tilde T$ varies much more gently with $1/\tilde a_R$ for states with $n > 40$ than the heuristic curve predicts (i.e., from Eq. \ref{ScaledT}, using Eqs. \ref{fVs1OvaRheuristic} and \ref{betaheur}), and there is no evidence for a plateau in the experimental result. While the ``end points'' of $(1/\tilde a_R, \, \tilde T) \approx (0.02, 0.7)$ and $\approx (0.12, 2.5)$ agree, the experimental $\tilde T$ values between these limits are significantly higher than those given by the heuristic. This difference seems to have two primary causes. First, as noted above, Eq. \ref{ScaledT} does not account for the kinetic energy of the Rydberg atoms. If this effect were to be included in Eq. \ref{ScaledT}, the $\tilde T$ values predicted would be higher in the range of $1/\tilde a_R$ where the final ionization fraction is significant. This is exactly the behavior we see in the data, as well as in many of the numerical simulations in Fig. \ref{thScaledTvsscaleda}. Second, as can be seen in Fig. \ref{thScTVsFf}, $\tilde T$ does not rise above a value of 1.6 (the $f_f \rightarrow 0$ limit) until $f_f \approx 0.5$. Using Fig. \ref{thFfVs1OvScaleda}(a), we see that this occurs for $f_i = 0.01$ at $1/\tilde a_R \approx 0.05$. If this onset occurs in the experiments at smaller $1/\tilde a_R$ than in the simulations, the rise in $\tilde T$ as $1/\tilde a_R$ increases would be more gradual than the heuristic curve. Assumptions made in the way the programs calculate the probabilities of the various different outcomes of each electron-atom collision could give rise to such a difference \cite{robx03}. For instance, the details of radiative cascades within the Rydberg ensemble are critically dependent on population remixing of different $n,\ell$ states due to electron-Rydberg collisions, and how each of these states decays radiatively \cite{robx17p}. 

We conclude from this analysis that the observed behavior of $\tilde T$ versus $1/\tilde a_R$ shown in Figs. \ref{scaledTvsscaleda} and \ref{thScaledTvsscaleda} for $n > 40$ (roughly, $1/\tilde a_R > 0.03$) is determined by ionizing electron-Rydberg collisions and is consistent with energy conservation as the plasma evolves. For $f_f \lesssim 0.5$, approximately 60\% of the initial Rydberg binding energy is converted to ion kinetic energy. Above $f_f = 0.5$, a regime approached in our experiments and simulations only for $n \ge 80$, the initial ionization mechanisms which seed the plasma lead to low $T_{e,i}$ plasmas which expand slowly, but which ultimately ionize as much as $\gtrsim 80$\% of the atoms. To conserve energy, the remaining Rydberg atoms are scattered to more deeply bound states, leading to the significant increase in $\beta$ above $f_f = 0.5$. This leads to the final ion kinetic energy increasing as a fraction of the initial Rydberg binding energy (i.e., an increase in $\tilde T$), since the energy released when a Rydberg atom is deexcited is proportional to its binding energy \cite{robx03}. 

With regard to the Coulomb coupling parameter for the electrons, $\Gamma_e$, using Eq. \ref{gammaScaled} and $\tilde T = 0.6$ gives $\Gamma_e = 1.6/\tilde a_e$. This coupling reaches its maximum value of $\Gamma_e \approx 0.1$ near $1/\tilde a_e \approx 1/\tilde a_R = 0.07$. However, at larger values of $1/\tilde a_R$, $\tilde T$ increases because the ionization fraction becomes large as described above, and $\Gamma_e$ decreases to $\approx 0.05$ for the highest Rydberg states we looked at, $n = 120$. Interestingly, this behavior is also well-described using the threshold lowering (TL) picture \cite{hahn01,hahn02}. In this regime, the atom cores are close enough such that $a_{R} \sim 2{n^\ast}^2a_0$, and the atom potential wells overlap. This lowers the ionization threshold by an amount $\Delta = 2 C_P \times e^2/4 \pi \epsilon_0 a_{R}$ (in SI units), where $C_P$ is a constant found to be $C_P = 11 \pm 5$ using a self-consistent calculation which accounts for the three-dimensional distribution of atom/ion cores. In this picture, if the Rydberg state lies within an energy $\Delta$ of isolated-atom ionization limit, laser excitation actually creates a free electron with temperature $T_{e,0}$ such that $\Delta = |E_{b,i}| + \frac{3}{2} k_B T_{e,0}$. The TL condition is thus $\tilde T = \frac{2}{3} (\frac {2C_P}{\tilde a_R} - 1)$, and this prediction for $\tilde T$ is shown in Fig. \ref{thScaledTvsscaleda} (d).

\subsection{{Results for $n \lesssim 40$}\label{templow}}

The numerical modeling approach also gives some insight into the evolution of Rydberg plasmas with $n \lesssim 40$. The results shown in Figs. \ref{thScaledTvsscaleda}(b) and (c) for $1/\tilde a_R < 0.02$ show quite good agreement with the experimental data in Fig. \ref{scaledTvsscaleda}. There is significant scatter in the experimental results, filling nearly the entire range between the lines corresponding to $\Gamma_e = 0.02$ to $0.05$, and the results of the numerical simulation are also in this range. However, closer inspection of Figs. \ref{thScaledTvsscaleda}(b) and (c) shows that there is a small but distinct systematic trend in the $\tilde T$ values. Specifically, for $T_{e,i} = 50$ K, the $\tilde T$ values are systematically higher than for $T_{e,i} = 5$ K, 10 K, and 25 K, for which the $\tilde T$ values are reasonably consistent. As can be seen in Fig. \ref{thFfVs1OvScaleda}(a), the difference between the $T_{e,i} = 50$ K and the 5, 10, and 25 K simulations is that the $f_f$ values are significantly higher for 50 K than for the other temperatures. Indeed, for 5, 10, and 25 K, the final ionization fraction is almost equal to the initial value used in the simulations, $f_f \approx f_i$. In this parameter range the avalanche is not a period when there is significant additional ionization. However, there is still significant exchange of energy between the plasma and the atoms during this period with the net result that the electrons are heated and the Rydberg atoms are deexcited. 

The behavior of the $T_{e,i} =$ 10 and 25 K results is to be expected in the regime where $n \le 40$. Basically, the initial electrons are too cool to ionize the parent Rydberg atoms because $|E_{b,i}| > 4 k_B T_{e,i}$. The $T_{e,i} = 50$ K simulations are the closest to the $\tilde T = 0.6$ line (i.e., the prediction of Eq. \ref{ScaledT} as $f_f \rightarrow 0$), and this supports this argument that the initial 50 K electrons can cause enough ionization that the UNP evolution loosely approximates to the mechanism described in Section \ref{temp40}, but for the lower $T_{e,i}$ electrons, this model is not valid. The fact that the experimental data in Fig. \ref{scaledTvsscaleda} parallel the simulations in Fig. \ref{thScaledTvsscaleda} for $n \lesssim 40$ ($1/\tilde a_R < 0.02$) for $T_{e,i} =$ 10 K and 25 K implies that, whatever the initial ionization mechanism in the experiments, it cannot produce electrons hot enough to cause ionization for samples with $n \lesssim 40$. The phenomenon is likely related to the decrease in ionization rates for all three plasma seeding processes as $n$ decreases, as well as the fact that the number of ions needed to reach threshold for plasma formation and the ion potential well depth are proportional to each other. In other words, low ionization rates means fewer ions, which produces a potential well that can only trap electrons which are too cool to ionize the parent atoms. 

Both the simulations in Fig.\,\ref{thScaledTvsscaleda} for low $T_{e,i}$, and the experimental results in Fig.\,\ref{scaledTvsscaleda}, show that the interaction between the plasma and cold Rydberg samples with $n \lesssim 40$ results in UNPs which evolve at constant $\Gamma_e$. However, the $\Gamma_e$ values for such plasmas are significantly smaller than we see at higher $n$. Specifically, the data in Fig. \ref{thScaledTvsscaleda} (b) and (c) ($f_i = 0.01$ and 0.1, respectively) fall approximately on the line $\Gamma_e = 0.03$. However, the lines in Fig. \ref{thScaledTvsscaleda} are drawn assuming that $\tilde a_e = \tilde a_R$. Using $\tilde a_e = \tilde a_R/f_f^{1/3}$ and $f_f = f_i$, the actual value is $\Gamma_e \approx 0.01$ for $1/\tilde a_R < 0.01$. In UNPs created by photoionization,  evolution at constant $\Gamma_e$ occurs due to competition between two limiting behaviors \cite{robx02}. First, in the absence of TBR, adiabatic expansion cools the electrons faster than the rate at which $a_e$ increases, and this would cause $\Gamma_e$ to increase with time. On the other hand, TBR both increases $T_e$ and reduces the electron density, thus increasing $a_R$, both of which would decrease $\Gamma_e$. For sufficiently high electron density and low electron temperatures, the balance between these two behaviors equilibrates the UNP to $\Gamma_e \sim 0.1$. 

For UNPs evolving from Rydberg atoms with $n \lesssim 40$ in which the electrons are too cold to cause ionization, the TBR heating mechanism is replaced with Rydberg deexcitation collisions, which become more probable than exciting collisions below the bottleneck energy. While the TBR rate is proportional to $\rho_e^2 \, T_e^{-9/2}$, the Rydberg deexcitation rate is proportional to $\rho_e \, T_e^{-0.17}$ \cite{mans69,robx03}. The deexcitation rate clearly has much weaker dependence on electron density and temperature than TBR, and additionally does not increase $a_e$. Based on this argument, deexcitation collisions have a weaker ability than TBR to heat the UNP and decrease $\Gamma_e$. On the other hand, some of this effect is offset by the fact that the amount by which deexcitation collisions heat the electrons is proportional to $|E_b|$ \cite{robx03}, and this is typically much greater in our situation for Rydberg plasmas with $n \lesssim 40$ than in photoionization-initiated UNPs where each TBR collision heats the plasma by an amount $\sim k_B T_e$. The heating provided by deexcitation collisions in this limit is sufficient to counterbalance the tendency for $\Gamma_e$ to increase due to adiabatic expansion, but the plasmas are limited to $\Gamma_e \sim 0.01$ because the initial ionization mechanisms cannot provide a high enough electron density for $\Gamma_e \sim 0.1$ to be reached with the typical Rydberg densities used in our experiments. 

The concept of avalanche ionization of Rydberg atoms by the UNP is not appropriate for Rydberg UNPs from atoms with $n \lesssim 40$ since no significant additional ionization happens after the plasma reaches threshold. In this regime, the ``avalanche'' is actually the time in which the plasma heats due to Rydberg deexcitation collisions, and this process does not have a well-defined end time. Rather, this electron heating mechanism gradually tapers off since the most of the Rydberg atoms get left behind as the plasma expands. The only Rydberg atoms ``carried along'' as the plasma expands are those in the ionization-recombination cycle, just as in a photoionization-initiated UNP. This will be a much smaller number than the parent Rydberg ensemble since only a small fraction of this sample has ionized before the avalanche begins. Additionally, the stationary parent ensemble will undergo radiative decay at higher rates as the initial $n$ decreases, further reducing its interaction with the UNP. For the $nd$ states used in this work, the radiative lifetimes are 35 $\mu$s at $n = 40$, but only 16 $\mu$s at $n = 30$ \cite{bran10}. Electron-Rydberg collisions will populate nearby high angular momentum states with much longer lifetimes, thereby mitigating some of this decline. However, there isn't enough time for the few electrons (relative to the number of parent Rydberg atoms) to populate anything close to a statistical ensemble of $\ell$ states, which would have effective lifetimes in the 100 $\mu$s - 1 ms range for $n = 30 - 40$ \cite{chan85}. We see the effect of the declining radiative lifetime at low $n$ reflected in the number of atoms which end up in states with $n \le 5$ at 40 $\mu$s of evolution time: at $n = 40$ and above, this number is negligible, but for $n \le 30$, the fraction in $n \le 5$ states is typically at least 50\%.  

\section{{CONCLUSION}\label{conc}}

We have described an experimental and numerical study of the effective initial electron temperatures, $T_{e,0}$, in ultra cold Rydberg plasmas. We find that, for plasmas which evolve from Rydberg samples with $n > 40$ in the density range $10^7 \, - \, 10^9$ cm$^{-3}$, the final ion kinetic energy, $(3/2) \, k_B \, T_{e,0}$, is related to the fraction of atoms which ionize. To maintain the energy balance, the remaining Rydberg atoms are much more deeply bound than the original state, and electron collisions with these atoms are more likely to heat the UNP than to cool it. In this regime, $T_{e,0}$ corresponds to between $0.6 \times |E_{b,f}|/k_B$ (for very low ionization fractions) and $2.5 \times |E_{b,f}|/k_B$ for the highest ionization level observed in our simulations, $f_f \approx 0.83$. Additionally, we find that $T_{e,0}$ is independent of the initial ionization mechanism which seeds the plasma. For cold Rydberg samples with $n \lesssim 40$, the initial ionization mechanisms which seed the plasma produce electrons which are too cold to cause further ionization. In this situation, the plasma evolves with constant, low, $\Gamma_e$ values due to competition between adiabatic cooling and electron-Rydberg collisions which deexcite the atoms and heat the electrons. 

\section{{ACKNOWLEDGEMENTS}\label{ack}}

We are deeply indebted to F. Robicheaux for sharing his programs and advice on how to run them, as well as extensive conversations at all stages of this project. In addition, we acknowledge extensive discussions with T. F. Gallagher and C. W. S. Conover, as well as the loan of equipment from the former. This work has been supported by Colby College through the Division of Natural Sciences grants program, by Middlebury College, and by NSF (Grant No. 1068191).



\raggedright
\bibliography{PRL_Bib_092116} 

\end{document}